# How Short Sales Circumvent the Capital Gains Tax System

## By Russell Stanley Q. Geronimo[*]

Through a short sale, a person borrows a share of stock from a lender, sells the borrowed share to a third person at the current price, and purchases an identical share in the market at a future date and at a future price to replace the borrowed share of stock.[1] This only makes sense if the short seller anticipates a downward trend in share price. The short seller incurs a gain if share price decreases because the cost of replacing the borrowed share falls below the selling price.[2] The reverse is true in an ordinary sale, where a person owning a share of stock incurs a loss if price decreases because the selling price falls below the basis or acquisition cost.[3]

Therefore, when a taxpayer simultaneously owns a share of stock and short sells an identical stock, any gain in an ordinary sale of the owned stock is offset by a corresponding loss in the short sale of the borrowed identical stock, *vice versa*.[4] This offsetting effect, in turn, creates an unexpected <u>tax deferral opportunity</u> abused in other jurisdictions[5] and which remains unregulated in the Philippine tax system.[6]

This tax deferral scheme is enabled by a short sale of a security identical to one already owned.[7] It gives the taxpayer an ability to create the economic equivalent of a disposition of a share of stock <u>without triggering a taxable realization event</u>.[8] We summarize the procedure and effect of this scheme as follows:

> Given a taxpayer who acquires x number of shares at time 1 and incurs unrealized gains at time 2, he can "cash out" or simulate the realization of these gains at time 2, and defer the payment of capital gains tax to time 3, if the taxpayer short sells x number of identical shares at time 2, and replaces the borrowed shares at time 3 using the shares acquired at time 1.[9]

---

[*] University of the Philippines - College of Law
[1] *Strategic Alliance Development Corp. vs. Radstock Securities Limited* (G.R. No. 178158, December 04, 2009); Section 2(r) of the *Rules on Securities Borrowing and Lending* (SEC Memorandum Circular No. 7 series of 2006); Section 135 of Revenue Regulation No. 02-40 dated February 10, 1940
[2] *White vs. Smith et. al.*, 54 N.Y. 522 (N.Y. 1874)
[3] Section 40(A) of the National Internal Revenue Code (NIRC) of 1997, as amended
[4] *U.S. v. Wood*, 364 F.3d 704 (6th Cir. 2004)
[5] Bray, Christopher P., *Estate Planning with Short Sales*, 74 TAXES 261 (1996); Frank, Mary Margaret, *Effective 'Estee-Te' Tax Planning Through Financial Engineering: Estee Lauder Companies Inc.*, Darden Case No. UVA-C-2261, available at SSRN: http://ssrn.com/abstract=1277000
[6] The U.S. regulatory framework is embodied in 26 U.S. Code § 1259 (Constructive sales treatment for appreciated financial positions). No similar rule exists in Philippine tax law.
[7] This is also called as a "short sale against the box". The word "box" refers to the traditional place of storage for stock certificates, which evidence shares owned by a stockholder. *See* Weisbach, David A., *Should a Short Sale Against the Box be a Realization Event?*, 50 NATIONAL TAX JOURNAL 3, 495 (1997)
[8] Whitmarsh, Theodore F., *When to Sell Securities Short Against the Box*, 28 FINANCIAL ANALYSTS JOURNAL 3, 80 (1972)
[9] *See* Dyl, Edward A., *Short Selling and the Capital Gains Tax*, 34 FINANCIAL ANALYSTS JOURNAL 2, 61 (1978); *Federal Taxation of Short Sales of Securities*, 56 HARVARD LAW REVIEW 2, 274–282 (1942); *Dubman v. North Shore Bank*, 85 Wis.2d 819 (Wis. Ct. App. 1978); *Bissell v. Merrill Lynch Co., Inc.*, 937 F. Supp. 237 (S.D.N.Y. 1996); *Reynolds v. Texas Gulf Sulphur Company*, 309 F. Supp. 548 (D. Utah 1970)

To elaborate, the taxpayer owns a share of stock at time 1 and short sells an identical stock at time 2.[10] But before selling the stock short, he must first borrow the stock, at which point he has two shares at hand as of time 2: the owned stock and the borrowed stock.[11] Under the existing tax realization regime, if he disposes the borrowed stock and claims the proceeds from the short selling transaction at time 2, he does not trigger a taxable realization event.[12] He only incurs a capital gains tax when he replaces the borrowed stock at time 3.[13]

Meanwhile, the taxpayer has already cashed out the proceeds of the short sale at time 2.[14] Since he owns a share of stock identical to the borrowed stock, any unrealized gain or loss from the owned stock is offset by an equivalent loss or gain from the short selling transaction, such that the taxpayer is already immune from the risk of price fluctuation between time 2 and time 3.[15]

At time 3, when he replaces the borrowed stock *using* the owned stock, he triggers two realization events: one pertains to the realization of gain or loss from the disposition of the owned stock, and another pertains to the realization of loss or gain from the short selling transaction upon replacement of the borrowed stock.[16] Because these two realization events offset each other financially, the taxpayer's net capital gains at time 3 is *equivalent* to what his taxable net capital gains would be had he sold the owned stock at time 2 without undertaking a short selling transaction.[17] Therefore, the taxpayer effectively reaps the economic benefits of selling the owned stock at time 2 and defers the payment of capital gains tax to time 3.[18]

To prevent this tax deferral scheme, we propose a new tax treatment of short sales if the taxpayer owns an identical security. Instead of having two realization events at time 3, the new tax rule should treat the short selling transaction at time 2 as a "constructive" disposition of the owned stock, which the taxpayer acquired at time 1, even though what the taxpayer sold was the borrowed identical stock.[19] This constructive disposition should trigger a taxable realization event at time 2 (i.e. when he entered the short sale) and another at time 3 (i.e. when he replaced the borrowed stock).[20] As will be shown in the foregoing discussions, this new tax treatment will result in (1) the elimination of the tax deferral opportunity created by short sales and (2) higher collection of capital gains tax from sale of securities.

Plugging the loophole in the existing realization rule is timely and necessary, considering that the Philippine Stock Exchange (PSE) is currently in the process of institutionalizing the short selling of stocks, starting with shares forming part of the PSE Index (PSEi).[21] The PSE[22], together with the Bangko Sentral ng Pilipinas (BSP)[23], the Bureau of Internal Revenue (BIR)[24], the Insurance Commission (IC)[25], and the Securities and Exchange Commission (SEC)[26], already laid down the legal and regulatory framework for short selling transactions in the securities market. In 2014, the PSE entered into a technology sales agreement for the purchase of a new trading system from Nasdaq OMX "as part of an overhaul to […] allow the widespread use of short selling."[27]

---

[10] Paul, Deborah L. *Another Uneasy Compromise: The Treatment of Hedging in a Realization Income Tax*. 3 FLA. TAX REV. 1 (1996)
[11] *Federal Taxation of Short Sales of Securities*, 56 Harvard Law Review 2 (1942)
[12] *Doyle v. Commissioner*, 286 F.2d 654, 657 (7th Cir. 1961)
[13] *Doyle v. Commissioner*, 286 F.2d 654, 657 (7th Cir. 1961)
[14] Hayward, Paul D., *Monetization, Realization, and Statutory Interpretation*, 51 Canadian Tax Journal 5, 1761 (2003)
[15] Whitmarsh, Theodore F., *When to Sell Securities Short Against the Box*, 28 Financial Analysts Journal 3, 80 (1972)
[16] Dyl, Edward A., *Short Selling and the Capital Gains Tax*, 34 Financial Analysts Journal 2, 61 (1978)
[17] *Supra* note 14.
[18] Weisbach, David A., *Should a Short Sale Against the Box be a Realization Event?*, 50 National Tax Journal 3, 495 (1997)
[19] *Id.*
[20] *Id.*

The absence of a special tax treatment for short sale of securities in any of these regulatory issuances creates a gap in law and regulation. Without a constructive disposition rule, as proposed herein, we would expose the capital gains tax system to millions of deferred or avoided capital gains tax.[28] In this article, we shall demonstrate how short sales create these tax deferral opportunities in the tax system. We shall also present a comprehensive analysis of how short sales generate a loophole in the realization rule. Lastly, we shall discuss the proposed tax treatment needed to close this gap in tax regulation.

Let us begin with an illustration. Consider a hypothetical investor named "X", who is not a dealer in securities[29], and a hypothetical security[30] called "ABC share of stock"[31], which is not traded in the stock exchange[32] and is a capital asset[33] of X. Let the price of ABC shares over a given time period be as follows:

---

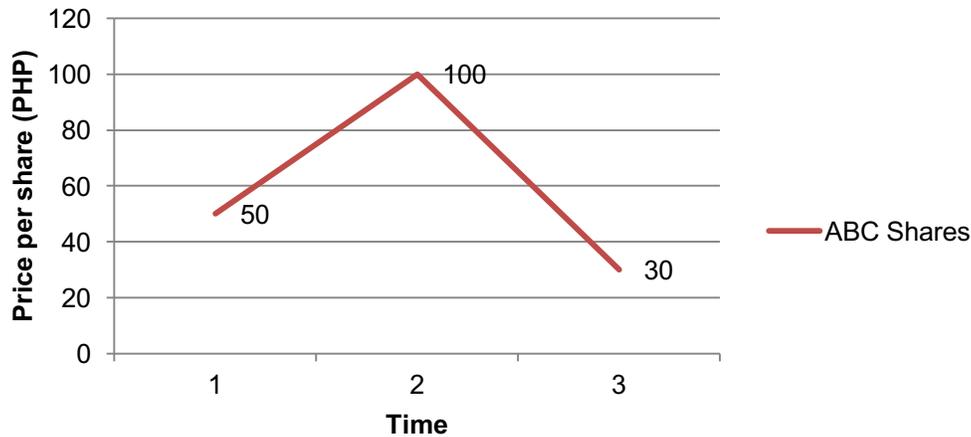

**ABC Share Price**
*Figure 1*

Next, we shall distinguish the tax implications of an ordinary sale and a short sale, with the objective of showing that the gain in an ordinary sale is offset by a loss in a short sale, *vice versa*.

TAX TREATMENT OF ORDINARY SALE OF STOCK

Suppose X purchases 100,000 ABC shares at time 1 and sells them at time 2.[34] X incurs a capital gain and he becomes liable for capital gains tax, as follows:

|  | Per Share (₱) | Total (₱) |
|---|---:|---:|
| Selling Price | 100 | 10,000,000 |
| Less: Basis | 50 | 5,000,000 |
| Capital Gain | 50 | 5,000,000 |
| Multiply by: Rate of Capital Gains Tax | 10% | 10% |
| Capital Gains Tax | 5 | 500,000 |

---

[32] *See* Section 24(C), Section 25(A)(3), Section 27(D)(2), Section 28 (A)(7)(C), and Section 28 (B)(5)(C) of the NIRC of 1997 for different tax treatments of capital gains from shares of stock not traded in the stock exchange.

[33] Section 39(A)(1) of the NIRC of 1997 states, "The term 'capital assets' means property held by the taxpayer (whether or not connected with his trade or business), but does not include stock in trade of the taxpayer or other property of a kind which would properly be included in the inventory of the taxpayer if on hand at the close of the taxable year, or property held by the taxpayer primarily for sale to customers in the ordinary course of his trade or business, or property used in the trade or business, of a character which is subject to the allowance for depreciation provided in Subsection (F) of Section 34; or real property used in trade or business of the taxpayer."

[34] An upward trend in share price is "bullish". (*State v. Plummer*, 117 N.H. 320, N.H. 1977)

The selling price of ₱100 per share is the amount realized from the sale. Section 40(A) of the NIRC states, "The amount realized from the sale or other disposition of property shall be the <u>sum of money received</u> […]".

The basis is the acquisition cost of ABC shares at time 1, which is ₱50 per share. Section 40(B)(1) of the NIRC states that the basis of the property sold shall be "the <u>cost</u> thereof in the case of property acquired on or after March 1, 1913, if such property was <u>acquired by purchase</u>".

The gain is ₱50 per share, being the difference between the selling price and the acquisition cost. Section 40(A) of the NIRC states, "The <u>gain</u> from the sale or other disposition of property shall be the <u>excess</u> of the <u>amount realized</u> therefrom over the <u>basis</u> or adjusted basis for determining gain".

We characterize the gain as a capital gain because ABC share is a capital asset of X, pursuant to the definition of a capital asset in Section 39(A)(1) of the NIRC.[35] Besides, we said that X is not a dealer in securities. Had he been a dealer in securities, the ABC share would be an ordinary asset, and the gain therefrom an ordinary income, pursuant to Section 22(Z) of the NIRC.[36]

The sale is a taxable realization event, pursuant to Section 40(C) of the NIRC, which states, "[…] upon the <u>sale</u> or exchange of property, the entire amount of the gain or loss, as the case may be, shall be recognized." Accordingly, the date of realization is time 2, which is the date of the sale.

Finally, we use a capital gains tax rate of 10% because ABC shares are not listed in the stock exchange, pursuant to Section 24(C) of the NIRC, which states:

> […] a final tax at the rates prescribed below is hereby imposed upon the net capital gains realized during the taxable year from the sale, barter, exchange or other disposition of shares of stock in a domestic corporation, except shares sold, or disposed of through the stock exchange:
>
> Not over P100,000 .......................................... 5%
> Amount in excess of P100,000 ........................ 10%

Now suppose that, instead of buying at time 1 and selling at time 2, X purchases 100,000 ABC shares at time 2 and sells them at time 3.[37] X incurs a capital loss, as follows:

---

[35] *Supra* note 14.
[36] The provision states, "The term 'ordinary income' includes any gain from the sale or exchange of property which is not a capital asset or property described in Section 39(A)(1). Any gain from the sale or exchange of property which is treated or considered, under other provisions of this Title, as 'ordinary income' shall be treated as gain from the sale or exchange of property which is not a capital asset as defined in Section 39(A)(1). The term 'ordinary loss' includes any loss from the sale or exchange of property which is not a capital asset. Any loss from the sale or exchange of property which is treated or considered, under other provisions of this Title, as 'ordinary loss' shall be treated as loss from the sale or exchange of property which is not a capital asset."
[37] A downward trend in share price is "bearish". (*Baviera vs. Paglinawan*, G.R. No. 168380, February 08, 2007)

|  | Per Share (₱) | Total (₱) |
|---|---|---|
| Selling Price | 30 | 3,000,000 |
| Less: Basis | 100 | 10,000,000 |
| Capital Loss | 70 | 7,000,000 |

Since the sale occurred at time 3, the capital loss was also realized at time 3, pursuant to Section 40(C) of the NIRC. And since the acquisition cost is higher than the selling price, X incurs a capital loss, pursuant to Section 40(A) of the NIRC, which states, "the loss shall be the <u>excess</u> of the <u>basis</u> or adjusted basis for determining loss over the <u>amount realized</u>".

X is not liable for capital gains tax because Section 24(C) of the NIRC states, "a final tax […] is hereby imposed upon the <u>net capital gains</u> realized". Absent other capital gains from other capital assets, X does not have a net capital gain, pursuant to Section 39(A)(2) of the NIRC, which states, "The term 'net capital gain' means the excess of the gains from sales or exchanges of capital assets over the losses from such sales or exchanges."

Now let us mirror these two transactions using a short sale, instead of an ordinary sale. But first, let us explain the legal nature of a short sale, and then we shall explain the tax treatment of a short sale of stock.

NATURE OF SHORT SALE

A short sale is an ordinary buy-and-sell transaction <u>in reverse sequence</u>,[38] enabled by a securities borrowing and lending agreement.[39] X, in effect, sells "what he does not have" because he borrows in order to sell.[40] This is implemented as follows:

(Step 1)   X borrows a share of stock from a lender,
(Step 2)   sells the borrowed share to a buyer, and
(Step 3)   buys an identical share from a seller, in order to
(Step 4)   replace the borrowed share to the lender.[41]

This contemplates two transactions: the securities borrowing and lending transaction, and the short sale proper. The securities borrowing and lending agreement is between the borrower and the lender of security, while the short sale proper is between the short seller and the buyer of security. The lender is not privy to the short sale proper, while the buyer in the short sale proper is not privy to the securities borrowing and lending transaction.

Section 2(r) of the *Rules on Securities Borrowing and Lending* (SEC Memorandum Circular No. 7 series of 2006) defines a securities borrowing and lending agreement, as follows:

---

[38] *CIR v. Ferree*, 84 F.2d 124 (3d Cir. 1936)
[39] See RULES ON SECURITIES BORROWING AND LENDING (SEC Memorandum Circular No. 7 series of 2006)
[40] *Supra* note 19.
[41] See note 75 of *Strategic Alliance Development Corp. vs. Radstock Securities Limited* (G.R. No. 178158, December 04, 2009)

> Securities Borrowing and Lending (SBL) means the lending of securities from a lender's portfolio on a given date to a borrower's portfolio to support the borrower's trading activities with the commitment of the borrower to return or deliver said securities or equivalent to the lender on a determined future date. This is also referred to as a Securities Lending Transaction (SLT).

The short sale has two important features. <u>First</u>, the object of sale must be <u>fungible</u>.[42] Section 2 of Revenue Regulation No. 1-2008 dated February 1, 2008 states:

> Being fungible in nature, the borrowed shares of stocks/securities are transferred from the Lender to the Borrower.

This enables the borrower to sell the borrowed share in step 2, obtain an identical share in step 3, and return the borrowed share to the lender in step 4.[43] The lender is usually a broker, dealer in securities, or any other financial institution.[44]

<u>Second</u>, notwithstanding the use of the term "borrow", the act of "borrowing" the fungible object in step 1 instantly transmits ownership from the lender to the borrower.[45] This makes the alienation from the borrower to a buyer valid in step 2.[46] Hence, the layman's notion that short selling is selling "what one does not have" can be quite misleading. The short seller already has title over the object of the sale before alienating the borrowed object; he merely has an obligation to return an identical object to the lender.[47]

Section 2(r) of the *Rules on Securities Borrowing and Lending* (SEC Memorandum Circular No. 7 series of 2006) recognizes the transfer of title over the security from lender to borrower:

> <u>Notwithstanding</u> the use of expressions such as "borrow", "lend", "loan", "return", "redeliver", in SBL transactions, <u>title to securities "borrowed" or "lent" shall pass from one party to another</u>, and the party obtaining such title is obligated to redeliver or return equivalent securities.

---

[42] "The quality of being fungible depends upon the possibility of the property, because of its nature or the will of the parties, being substituted by others of the same kind, not having a distinct individuality." (*BPI Family Bank vs. Amado Franco*, G.R. No. 123498, November 23, 2007)

[43] Section 2 of Revenue Regulation No. 1-2008 dated February 1, 2008 states, "Upon demand of the Lender or at the end of the stipulated borrowing period, the Borrower is then obligated to return the equivalent shares of stock/securities and the Lender, in turn, returns the collateral put up by the Borrower."

[44] Section 4 of Revenue Regulation No. 1-2008 dated February 1, 2008 states, "A Lender is any person, whether natural or juridical, who lends shares of stock/securities from his/its pool of assets. There are no restrictions on the status and qualifications of a person who enters into an MSLA as a Lender. A foreign lender is contemplated within the definition of a Lender for the purpose of these Regulations."

[45] Article 1953 of New Civil Code states, "A person who receives a loan of money or any other <u>fungible</u> thing acquires the ownership thereof, and is bound to pay to the creditor an equal amount of the same kind and quality."

[46] *See*, for e.g., *People vs. Teresita Puig* (G.R. No. 173654-765, August 28, 2008), applying Article 1953 of the New Civil Code in bank operations.

[47] "Being the owner, the borrower [of a fungible thing] can dispose of the thing borrowed and his act will not be considered misappropriation thereof." (*Tanzo vs. Drilon*, G.R. No. 106671, March 30, 2000)

The Separate (Concurring) Opinion in *Hemedes vs. Court of Appeals* (G.R. No. 107132, October 08, 1999) appears to describe a short sale as a sale of future things[48], as follows:

> […] the law does not prohibit but, in fact, sanctions the perfection of a sale by a non-owner, such as the sale of future things or a <u>short sale</u>, for it is only at the consummation stage of the sale, i.e., delivery of the thing sold, that ownership would be deemed transmitted to the buyer.

The statement reflects an antediluvian notion of short sales and is no longer in keeping with widespread commercial practices. A contract of sale has a perfection stage and a consummation stage.[49] According to the Opinion, in the perfection stage, the seller need not be an owner of the object of the sale. However, in the consummation stage, the seller must be an owner because he becomes obligated to deliver and transmit ownership to the buyer. This is because of the well-established rule that the owner alone has a right to transmit his ownership to another.[50] Therefore, the Opinion states that a short sale is valid because, even though the seller is a non-owner, the short sale occurs in the perfection stage, with a promise to deliver the object in the future, at which point the seller is already the owner and is capable of transmitting his ownership.[51]

*On the contrary*, a short sale is <u>not</u> necessarily[52] a sale of future things. A short seller is already the <u>owner</u> of a borrowed fungible object even in the perfection stage.[53] Therefore, he does not need to wait for the arrival of a <u>future</u> time to obtain ownership because he has title over the object at <u>present</u> time and is now ready to deliver it to the buyer.[54] Let us emphasize that the notion of a short sale being the sale of "what one does not have" is a layman's notion, and has no basis in law.

Footnote 75 of *Strategic Alliance Development Corp. vs. Radstock Securities Limited* (G.R. No. 178158, December 04, 2009) describes the nature of a short sale more accurately, as follows:

> Article 1459 of the Civil Code provides: "The thing must be licit and the vendor must have a right to transfer the ownership thereof at the time it is delivered." The vendor cannot transfer ownership of the thing if he does not own the thing or own rights of ownership to the thing. The only possible <u>exception</u> is in a <u>short sale</u> of securities or commodities, where the seller borrows from the broker or third party the securities or commodities the ownership of which is <u>immediately transferred</u> to the buyer. This is feasible only when the subject matter of the transaction is a <u>fungible</u> object.

---

[48] Article 1462 of the New Civil Code states, "The goods which form the subject of a contract of sale may be either existing goods, owned or possessed by the seller, or goods to be manufactured, raised, or acquired by the seller after the perfection of the contract of sale, in this Title called 'future goods.'"
[49] *Lim, Jr. vs. San*, G.R. No. 159723, September 09, 2004
[50] "If the title did not reside in the person holding the property […], his alienation thereof would necessarily be null and void, as executed without a right to do so and without a right which he could transmit to the acquirer." (*Edroso vs. Sablan*, G.R. No. 6878, September 13, 1913)
[51] *Hemedes vs. Court of Appeals*, G.R. No. 107132, October 08, 1999
[52] We say "not necessarily" because a short sale can be a sale of future things if so designed in the contract.
[53] Section 2(r) of the *Rules on Securities Borrowing and Lending* (SEC Memorandum Circular No. 7 series of 2006)
[54] Section 2 of Revenue Regulation No. 1-2008 dated February 1, 2008

Now we are ready to discuss the tax treatment of a short sale of stock.

TAX TREATMENT OF SHORT SALE OF STOCK

Going back to <u>Figure 1</u>, suppose that X borrows 100,000 ABC shares from a stock lender and immediately sells the borrowed shares to a third party at time 1. At time 2, he replaces the same number of ABC shares by purchasing 100,000 ABC shares from the market and by delivering these shares to the lender. X incurs capital loss, as follows:

|  | Per Share (₱) | Total (₱) |
|---|---|---|
| Selling Price from Short Sale | 50 | 5,000,000 |
| Less: Cost of Replacing Borrowed Shares | 100 | 10,000,000 |
| Capital Loss | 50 | 5,000,000 |

Note that the sale took place at time 1, while the purchase took place at time 2. Therefore, the selling price is the price of ABC shares at time 1, and the cost of replacing the borrowed shares is the price of ABC shares at time 2.[55] This is the complete reversal of the ordinary sale of stock.

The tax treatment of short sale of stock is provided in Section 39(F)(1) of the NIRC, which states:

> Gains or losses from short sales of property shall be considered as gains or losses from sales or exchanges of capital assets[.]

The computation of a capital gain or loss from a short sale, therefore, requires the analogous application of Section 40(A) of the NIRC for the determination of "amount realized", Section 40(B)(1) of the NIRC for the determination of "basis" of property sold, Section 40(A) of the NIRC for the determination of "gain" or "loss", and the various provisions of the NIRC prescribing the "rates" of capital gains tax.

The selling price of ₱50 per share is the amount realized from the short sale at time 1, pursuant to Section 40(A) of the NIRC. The basis is the cost of replacing the borrowed ABC shares at time 2, which is ₱100 per share, pursuant to Section 40(B)(1) of the NIRC. The loss is ₱50 per share, being the difference between the cost of replacing the borrowed shares and the selling price, pursuant to Section 40(A) of the NIRC.

Had X implemented an ordinary sale between time 1 and time 2 instead of a short sale, he would have generated a gain <u>of the same amount</u> as his loss in the short sale,[56] and he would have been liable for capital gains tax. This inverse relationship is illustrated as follows:

---

[55] *See,* e.g., *Lagrange v. CIR*, 26 T.C. 191, T.C. 1956
[56] This offsetting relationship is the basis for "hedging transactions". *See*, e.g., *Steward Silk Corp. v. CIR*, 9 T.C. 174, 1947; *Fed. Natl. Mortgage Ass'n v. CIR*, 100 T.C. 541, 1993; *Int'l. Flavors Fragrances v. CIR*,

| Ordinary Sale | | Short Sale | |
|---|---|---|---|
| | Per Share (₱) | | Per Share (₱) |
| Amount Realized (time 2) | 100 | Amount Realized (time 1) | 50 |
| Less: Basis (time 1) | 50 | Less: Basis (time 2) | 100 |
| Capital Gain (time 2) | 50 | Capital Loss (time 2) | 50 |
| Multiply by: Rate of CGT | 10% | | |
| Capital Gains Tax | 5 | | |

Now suppose that X borrows 100,000 ABC shares from a lender and immediately sells the borrowed shares to a third party at time 2. He replaces the same number of ABC shares to the lender by buying 100,000 ABC shares in the market at time 3. X incurs capital gains and capital gains tax, as follows:

| | Per Share (₱) | Total (₱) |
|---|---|---|
| Selling Price from Short Sale (time 2) | 100 | 10,000,000 |
| Less: Cost of Replacing Borrowed Shares (time 3) | 30 | 3,000,000 |
| Capital Gain (time 3) | 70 | 7,000,000 |
| Multiply by: Rate of Capital Gains Tax | 10% | 10% |
| Capital Gains Tax | 7 | 700,000 |

One critical question is, "when is X liable for capital gains tax?" And the answer, of course, depends on the timing of the realization event. The real question, therefore, is, "when did X realize the capital gain?"

In an ordinary sale, the taxable realization event is the sale, pursuant to Section 40(C) of the NIRC, which states, "[…] upon the sale or exchange of property, the entire amount of the gain or loss, as the case may be, shall be recognized." Accordingly, the date of realization in an ordinary sale is the date of the sale.[57] This is pursuant to the longstanding income tax principle that capital gains are recognized when they are realized, and they are realized when capital assets are sold, transferred, exchanged or disposed.[58]

The situation is different in a short sale. Here, the traditional buy-and-sell sequence is reversed: X sold the shares first, and then he subsequently purchased identical shares.[59] Of course, at the time that he sold the shares at time 2, he did not yet know the basis of the shares. It was only at time 3, when he bought shares to replace the borrowed shares, did X determine the cost of replacing the borrowed shares, and therefore the basis of the stock.[60]

---

62 T.C. 232, 1974; *Vickers v. CIR*, 80 T.C. 394, 1983; *Volkart Brothers, Inc. v. Freeman*, 311 F.2d, 5th Cir. 1962; and *John A. Franks Co. v. Bridges*, 337 Mass. 287, 1958

[57] "[A]n exchange of property gives rise to a realization event[.]" (*Cottage Savings Assn v. CIR*, 499 U.S. 554, 1991)

[58] This requirement was adopted from the U.S. income tax system and which originated from the Supreme Court ruling in *Eisner vs. Macomber* (252 U.S. 189, 1920), where it was held that a taxable gain must be derived and severed from capital. The *Eisner* doctrine was applied domestically in *CIR vs. A. Soriano Corp.*, G.R. No. 108576 January 20, 1999.

[59] *Supra* note 19.

Section 135 of Revenue Regulation No. 02-40 dated February 10, 1940, which implements Section 39(F)(1) of the NIRC, states:

> […] the short sale is <u>not deemed to be consummated</u> until the obligation of the seller created by the short sale is finally discharged by <u>delivery</u> of property to the brokers to replace the property borrowed […]

Since a short sale is consummated by the delivery of the object of sale <u>to replace the borrowed object</u>, the realization event occurred at time 3, and not at time 2. At time 2, X <u>received</u> the proceeds of the short sale. However, this is only a <u>receipt</u> of sale proceeds, and not a <u>realization</u> of gain.[61] At time 3, X delivered the borrowed share to the lender, and it was only at this point that the <u>receipt</u> made at time 2 was determined to include a gain. This is because the basis or cost of replacement was only determined at time 3.

In short sales, capital gains are realized upon the replacement of the borrowed stock, and not upon its disposition by the short seller.[62] This was adopted from Treasury Regulation 118, 39,117(g)-1 (1953), which states:

> For income tax purposes, a short sale is not deemed to be consummated until delivery of property to cover the short sale[.]

This special realization rule was upheld in *Doyle v. Commissioner*[63], which states that a "short sale is completed on the date the sale is covered, not at the time the order for the sale was entered into." In our illustration, the short sale was entered at time 2, and it was covered at time 3. By "covered", we mean that the obligation to return the borrowed stock has been complied with.

The rationale for this special realization rule is the same as and is consistent with the rationale for the realization rule in ordinary sale, as laid down in *Eisner vs. Macomber*[64], which states that gain must be derived and severed from capital. In entering the short sale, X in effect still does not know his invested capital. Hence, both the taxpayer and the tax collector cannot reckon the severance of gain from an undetermined invested capital.

*Eisner* led to *Commissioner vs. Glenshaw Glass Co.*[65], which states that a gain is an "undeniable accession to wealth, clearly realized, and over which the taxpayer has complete dominion." In receiving the proceeds of the short sale, the taxpayer does <u>not</u> have an *undeniable* accession to wealth. After all, he can still incur a loss when the short sale transaction is closed at a future date, when the price of stock increases above the proceeds he had received.

---

[60] *See Doyle v. CIR*, 286 F.2d 654, 7th Cir. 1961
[61] *See Filipinas Synthetic Fiber Corp. vs. Court of Appeals* (G.R. Nos. 118498 & 124377, October 12, 1999) for a distinction between realization and receipt in tax accounting.
[62] Section 135 of Revenue Regulation No. 02-40 dated February 10, 1940
[63] 286 F.2d 654, 657 (7th Cir. 1961)
[64] 252 U.S. 189 (1920)
[65] 348 U.S. 426, 431 (1955)

In entering the short sale transaction, and prior to returning the borrowed stock, the short seller is exposed to variation in the value of the stock, which he is obliged to return.[66] Hence, he is as much exposed to routine ups and downs of the marketplace as a person who continues to hold a stock.[67]

Considering the applicable realization rules for ordinary sale and short sale, compare their tax implications in the period between time 2 and time 3:

| Ordinary Sale | | Short Sale | |
|---|---:|---|---:|
| | Per Share (₱) | | Per Share (₱) |
| Amount Realized (time 3) | 30 | Amount Realized (time 2) | 100 |
| Less: Basis (time 2) | 100 | Less: Basis (time 3) | 30 |
| Capital Loss | 70 | Capital Gain (time 3) | 70 |
| | | Multiply by: Rate of CGT | 10% |
| | | Capital Gains Tax | 7 |

This emphasizes the perfect inverse relationship between the gains or losses in an ordinary sale and short sale, and is elaborated further in the next section.

OFFSETTING EFFECT OF SHORT SALE OF STOCK

One important relationship between an ordinary sale and a short sale is that, *given the same transaction period and involving identical securities*, the gain in an ordinary sale is a loss in a short sale, and the loss in an ordinary sale is a gain in a short sale.[68]

Under an ordinary sale, X already knows the basis of the share beforehand and only learns the selling price afterward, when he is about to dispose the share.[69] Under a short sale, X knows the selling price beforehand and only learns the basis afterward, when he is about to replace the borrowed share.[70] This counterintuitive sequence of events in a short sale is only consistent with the fact that the sale occurs first before the purchase.[71] The following is a summary of the distinctions between ordinary sale and short sale:

---

[66] *Cottage Say. Ass'n v. Commissioner*, 499 U.S. 554, 569-70 (1991)
[67] *Cottage Say. Ass'n v. Commissioner*, 499 U.S. 554, 569-70 (1991)
[68] Miller, David S. and Bertrand, Jean Marie, *The U.S. Federal Income Tax Treatment of Hedge Funds, Their Investors and Their Managers* (February 9, 2011), available at SSRN: http://ssrn.com/abstract=1758748
[69] Section 40(B)(1) of the NIRC states that the basis of the property sold shall be "the cost thereof in the case of property acquired on or after March 1, 1913, if such property was acquired by purchase".
[70] *Supra* note 39.
[71] *Supra* note 19.

|  | **Ordinary Sale** | **Short Sale** |
|---|---|---|
| **Transaction sequence** | Purchase first, dispose afterward[72] | Dispose first, purchase afterward[73] |
| **Related contracts** | Contract of sale[74] | Securities borrowing agreement[75] and contract of sale[76] |
| **Effect of increase in price of security** | Incur unrealized gain *after purchase* (i.e. while holding the security before disposition)[77] | Incur unrealized loss *after short sale but before replacement*[78] |
| **Effect of decrease in price of security** | Incur unrealized loss *after purchase* (i.e. while holding the security before disposition)[79] | Incur unrealized gain *after short sale but before replacement*[80] |
| **Knowledge of *basis* and proceeds of disposition** | *Basis* is known first, then amount of disposition proceeds is learned afterward[81] | Amount of disposition proceeds is known first, then *basis* is learned afterward[82] |

To illustrate the inverse relationship between the two transactions, consider two time periods: the present and the future. X is perfectly informed of the present price of ABC shares at ₱100 per share. However, X needs to speculate the future price of ABC shares, which can range from a low of ₱25 per share to a high of ₱175 per share, as follows:

| **Present Share Price (₱)** | **Probable Share Price in the Future (₱)** |
|---|---|
| 100 | 25 |
| 100 | 50 |
| 100 | 75 |
| 100 | 100 |
| 100 | 125 |
| 100 | 150 |
| 100 | 175 |

Next, consider implementing both an ordinary sale and a short sale, whereby X buys at the current date and sells at a future date under the ordinary sale, and X sells at the current date and buys at a future date under the short sale. This gives us the following table of possible gains or losses:

---

[72] *Supra* note 43.
[73] *Supra* note 19.
[74] Article 1458 of the New Civil Code
[75] Section 2(r) of the *Rules on Securities Borrowing and Lending* (SEC Memorandum Circular No. 7 series of 2006)
[76] Article 1458 of the New Civil Code
[77] Weisbach, D.A., *Short a Short Sale Against the Box Be a Realization Event?* 50 NATIONAL TAX JOURNAL 3, 495–506 (1997).
[78] *Id.*
[79] *Id.*
[80] *Id.*
[81] *Supra* note 43.
[82] *Supra* note 39.

| Ordinary Sale (₱) | | | Short Sale (₱) | | |
|---|---|---|---|---|---|
| Selling Price<br>Price at future date | Basis<br>Price at present date | Gain (Loss) | Selling Price<br>Price at present date | Basis<br>Price at future date | Gain (Loss) |
| 25 | 100 | -75 | 100 | 25 | 75 |
| 50 | 100 | -50 | 100 | 50 | 50 |
| 75 | 100 | -25 | 100 | 75 | 25 |
| 100 | 100 | 0 | 100 | 100 | 0 |
| 125 | 100 | 25 | 100 | 125 | -25 |
| 150 | 100 | 50 | 100 | 150 | -50 |
| 175 | 100 | 75 | 100 | 175 | -75 |

Note that the <u>selling price under the ordinary sale is equivalent to the basis in the short sale</u>, and the <u>basis in the ordinary sale is equivalent to the selling price in the short sale</u>. Because of the reverse sequence of events and inverse relationship, the gains and losses cancel each other out, as follows:

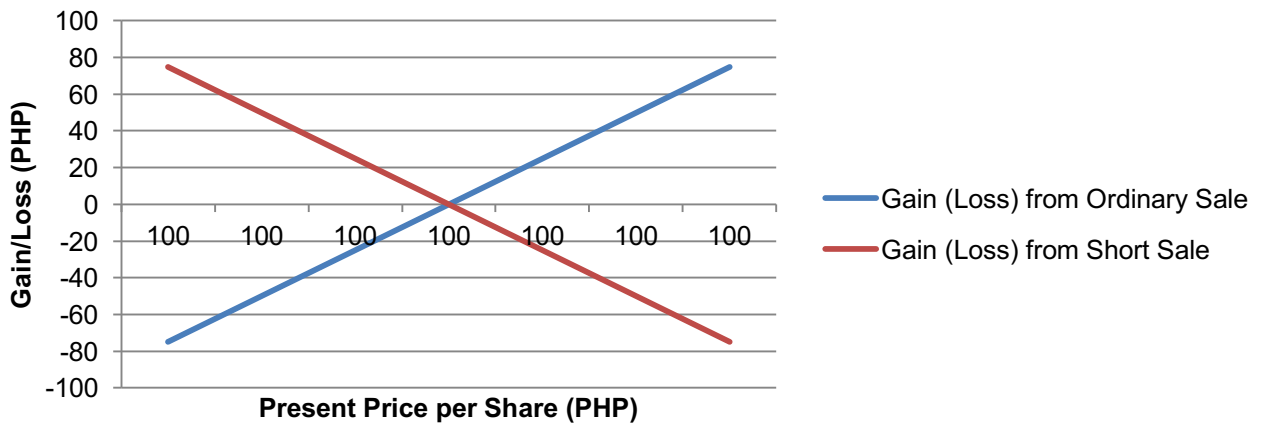

This offsetting effect is an important enabler of the tax deferral scheme to be discussed in the succeeding sections. Another important enabler is the timing difference in receipt of sale proceeds, as discussed in the next section.

TIMING DIFFERENCE IN RECEIPT OF SALE PROCEEDS

Another important distinction between an ordinary sale and a short sale is the timing of receipt of sale proceeds.[83] In an ordinary sale, where X buys at time 1 and sells at time 2, the date of receipt of sale proceeds is the date of realization, which is the

---

[83] Miller, David S., *Taxpayers' Ability to Avoid Tax Ownership: Current Law and Future Prospects*, 51 The Tax Lawyer 2, 279-349 (1998).

date of sale, and which is time 2.[84] In a short sale, where X sells at time 1 and buys at time 2, the date of receipt of sale proceeds is time 1, i.e. <u>before</u> the date of realization, which is the date when X replaces the borrowed shares at time 2.[85] This timing difference is illustrated as follows:

|  | Ordinary Sale | | Short Sale | |
|---|---|---|---|---|
| **Time period** | time 1 | time 2 | time 1 | time 2 |
| **Sequence of events** | Acquire (buy) | Dispose (sell) | Dispose (sell) | Acquire (buy) |
| **Date of realization** |  | ✔ |  | ✔ |
| **Date of receipt of sale proceeds** |  | ✔ | ✔ |  |

In the ordinary sale, X cashes out the benefits of sale only during the realization event.[86] In the short sale, X cashes out the benefits of sale prior to the realization event.[87] This advanced receipt of sale proceeds is the mechanism by which the taxpayer "cashes out" unrealized gains incurred as of the date of short sale (i.e. time 1) without triggering a taxable realization event, which is postponed to time 2.[88]

The difference between receipt and realization is not very obvious in an ordinary sale because the receipt happens contemporaneously with realization. In a short sale, X "receives" an amount which is not yet determined to be a "gain" because there is no determination yet of a basis.[89] There is only a gain when both "amount realized" and "basis" are determined, and there is an excess of amount realized over the basis.[90] This only happens at time 2.

REPLACING BORROWED SHARES WITH SHARES ALREADY OWNED

In a short sale, the borrower is obligated to replace the borrowed shares to the lender.[91] The presumption is that the borrower <u>purchases</u> identical shares in the market, and therefore the cost of replacing the borrowed share (which is analogous to an acquisition cost) is the prevailing share price.[92] Recall the following steps in a short sale of stock:

---

[84] Section 40(C) of the NIRC states, "[…] upon the <u>sale</u> or exchange of property, the entire amount of the gain or loss, as the case may be, shall be recognized."
[85] Hayward, Paul D., *Monetization, Realization, and Statutory Interpretation*, 51 CANADIAN TAX JOURNAL 5, 1761 (2003)
[86] Sherbaniuk, D.J., *Receipt and the Time of Recognition of Income: A Historical Conspectus of the Income Tax Laws of the United Kingdom, the United States and Canada*, 15 University of Toronto Law Journal 1, 62-101 (1963)
[87] *Supra* note 59.
[88] Schizer, David M., *Frictions as a Constraint on Tax Planning*, 101 COLUMBIA LAW REVIEW 6, 1312-1409 (2001)
[89] *Supra* note 59.
[90] *Id.*
[91] Duffie, Darrell, Nicolae Garleanu, and Lasse Heje Pedersen, *Securities Lending, Shorting, and Pricing*, 66 JOURNAL OF FINANCIAL ECONOMICS 2 307-339 (2002)
[92] D'avolio, Gene, *The Market for Borrowing Stock*, 66 JOURNAL OF FINANCIAL ECONOMICS 2, 271-306 (2002)

(Step 1)   X borrows a share of stock from a lender,
(Step 2)   sells the borrowed share to a buyer, and
(Step 3)   buys an identical share from a seller, in order to
(Step 4)   replace the borrowed share to the lender.[93]

On the other hand, it is entirely possible that X <u>already owns</u> shares identical to the one he borrowed in step 1.[94] In this case, he need not purchase shares in the market in step 3 to replace the borrowed shares in step 4. He simply delivers the shares that he already owns to the borrower.[95] We can therefore dispose of the need to buy identical shares for the purpose of replacing the borrowed shares. The new sequence of steps is as follows:

(Step 1)   X <u>acquires</u> ABC share.
(Step 2)   X borrows an <u>additional</u> ABC share from a lender. He now has two ABC shares at hand: the <u>owned</u> ABC share and the <u>borrowed</u> ABC share.
(Step 3)   X sells the <u>borrowed</u> ABC share to a third party buyer.
(Step 4)   X replaces the <u>borrowed</u> ABC share *by delivering the <u>owned</u> ABC share* to the lender.

In the <u>first scenario</u>, X does not own ABC shares. In the <u>second scenario</u>, X acquires ABC shares and still engages in a short sale of identical shares. In the <u>first scenario</u>, there is only <u>one</u> realization event, while in the <u>second scenario</u>, there are <u>two</u> realization events.[96]

In the <u>first scenario</u>, X realizes the gain or loss <u>from the short sale</u> at the time he replaces the ABC share to the lender. This is the <u>only</u> realization event. In the <u>second scenario</u>, X realizes the gain or loss <u>from the short sale</u> at the time he replaces the ABC share to the lender (the 1$^{st}$ realization event), <u>and also realizes the gain or loss *from disposing the owned ABC share*</u> (the 2$^{nd}$ realization event).[97]

In the <u>second scenario</u>, the disposition of the owned ABC share to replace the borrowed share is *analogous* to an ordinary sale. This is because X had to purchase an ABC share from the market <u>if he did not own an ABC share to begin with</u>. Since he owns an ABC share, it is <u>as if</u> he first sold the owned share in the market and used the sale proceeds to buy an identical share to replace the borrowed share. But since this requires an unnecessary step, he just delivers the owned share to the lender in order to replace the borrowed share. Accordingly, this delivery or transfer amounts to a realization event, <u>in addition to</u> the realization event created by the replacement of the borrowed share.[98]

An <u>ordinary sale</u> and the <u>disposition of the owned share to replace the borrowed share</u> are economically equivalent transactions, illustrated as follows:

---

[93] *Supra* note 22.
[94] Ulcickas, Simon D., *Internal Revenue Code Section 1259: A Legitimate Foundation for Taxing Short Sales Against the Box or a Mere Makeover*, 39 WM. & MARY L. REV. 1355 (1997)
[95] Feld, Alan L., *When Fungible Portfolio Assets Meet: A Problem of Tax Recognition*, THE TAX LAWYER 409-443 (1991)
[96] *Id.*
[97] *Supra* note 59.
[98] *Supra* note 68.

|  | Ordinary Sale (₱) | Disposition of Owned Share to Replace the Borrowed Share (₱) |
|---|---|---|
| Proceeds of Disposition | 100 | 100 |
| Less: Basis | 50 | 50 |
| Capital Gain | 50 | 50 |

Using the same price information in Figure 1, the capital gain from the disposition of <u>owned</u> ABC share is as follows:

|  | Per Share (₱) | Total (₱) |
|---|---|---|
| Proceeds of Disposition (time 2) | 100 | 10,000,000 |
| Less: Basis (time 1) | 50 | 5,000,000 |
| Capital Gain | 50 | 5,000,000 |

Meanwhile, the capital loss from the short sale is as follows, using the same price information in Figure 1:

|  | Per Share (₱) | Total (₱) |
|---|---|---|
| Selling Price from Short Sale (time 1) | 50 | 5,000,000 |
| Less: Cost of Replacing Borrowed Shares (time 2) | 100 | 10,000,000 |
| Capital Loss | 50 | 5,000,000 |

Accordingly, the net capital gain is <u>zero</u>, illustrated as follows:

|  | Per Share (₱) | Total (₱) |
|---|---|---|
| Capital Gain from Disposition of Owned Share | 50 | 5,000,000 |
| Less: Capital Loss from Short Sale | 50 | 5,000,000 |
| Net Capital Gain | 0 | 0 |

The zero net capital gain is a consequence of the fact that the loss in the short sale offsets the gain in the ordinary sale.[99]

TAX DEFERRAL SCHEME

We are now ready to illustrate how short selling a stock creates a tax deferral scheme by circumventing the realization rule.[100] Recall the information on ABC share prices in Figure 1, reproduced as follows:

---

[99] Miller, Mark A, *Hedging Strategies for Protecting Appreciation in Securities and Portfolios*, 15 JOURNAL OF FINANCIAL PLANNING - DENVER 8, 64-77 (2002)
[100] Knoll, Michael S, *Put-Call Parity and the Law*, 24 CARDOZO L. REV. 61 (2002)

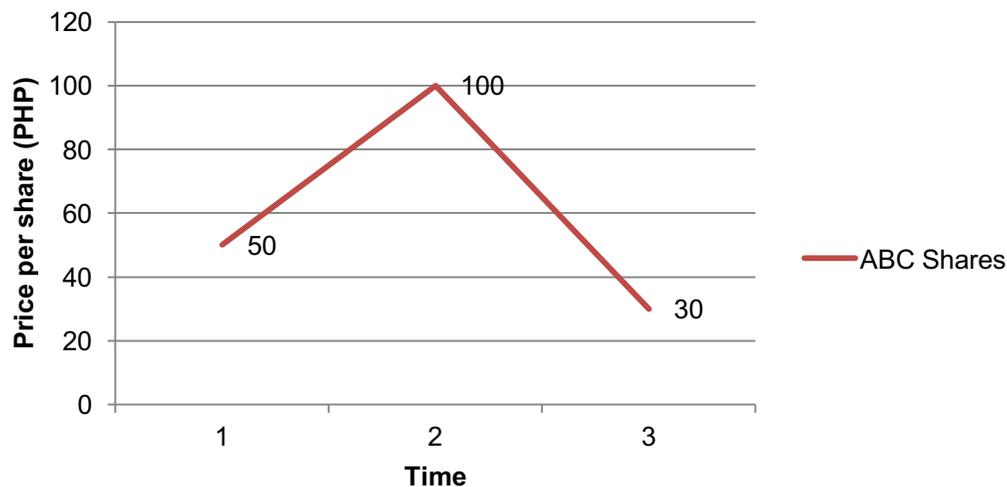

**ABC Share Price**
*Figure 1*

Suppose that time 1 pertains to a <u>past event</u>, time 2 to the <u>current date</u>, and time 3 to a <u>future date</u>. At <u>time 1</u>, X purchased an ABC share at ₱50 per share. At <u>time 2</u>, X has unrealized gain of ₱50 per share because of the increase in the price of ABC shares from ₱50 to ₱100 per share. At <u>time 3</u>, share price drops to ₱30 per share, wiping out the unrealized capital gain of ₱50 per share at time 2 and incurring an unrealized capital loss of ₱20 per share since time 1. However, because time 3 is a future event, the price of ABC shares at time 3 is <u>not yet known</u> to X and is only known when time 3 arrives.

At <u>time 2</u> (the current date), X is contemplating <u>three</u> strategic options. The <u>first strategy</u> is to sell the share at time 2 in order to immediately realize the gain of ₱50 per share. Since sale is a taxable realization event, it also entails the accrual of capital gains tax at time 2. This is illustrated as follows:

*STRATEGY 1*

|  | Per Share (₱) | Total (₱) |
|---|---|---|
| Original Purchase Price (time 1) | 50 | 5,000,000.00 |
| Add: Unrealized Capital Gains (time 1-2) | 50 | 5,000,000.00 |
| Share Price (time 2) | 100 | 10,000,000.00 |
|  |  |  |
| Selling Price (time 2) | 100 | 10,000,000.00 |
| Less: Basis (time 1) | 50 | 5,000,000.00 |
| Capital Gains (time 2) | 50 | 5,000,000.00 |
| Multiply by: Rate of Capital Gains Tax | 10% | 10% |
| Capital Gains Tax (time 2) | 5 | 500,000.00 |

The <u>second strategy</u> is to hold the share until time 3 in order to benefit from a <u>possible improvement</u> in capital gain, but this also entails the <u>probability of capital loss</u>. The capital gains tax depends on whether X makes a capital gain or capital loss at time

3. Suppose that X implements this strategy, and learns that the share price drops to ₱30 per share. X incurs a capital loss, illustrated as follows:

### STRATEGY 2

|  | Per Share (₱) | Total (₱) |
|---|---|---|
| Original Purchase Price (time 1) | 50 | 5,000,000.00 |
| Less: Unrealized Capital Loss (time 1-3) | 20 | 2,000,000.00 |
| Share Price (time 3) | 30 | 3,000,000.00 |
|  |  |  |
| Selling Price (time 3) | 30 | 3,000,000.00 |
| Less: Basis | 50 | 5,000,000.00 |
| Capital Loss (time 3) | (20) | (2,000,000.00) |

The <u>third strategy</u> is to implement a tax deferral scheme, where X can "cash out" the unrealized gain of ₱50 per share at time 2, but defer the payment of capital gains tax to time 3.[101] In order to defer the payment of capital gains tax, he must necessarily postpone the execution of a sale, which is the realization event, to time 3.[102] The detailed procedure for this strategy will be discussed shortly.

The three strategic options are summarized as follows:

|  | **Strategy 1** (Sell at time 2, the current date) | **Strategy 2** (Sell at time 3, the future date) | **Strategy 3** (Tax Deferral Scheme) |
|---|---|---|---|
| **Economic implication** | Realize gains immediately | Possibility of improving capital gain and risk of incurring capital loss | Cash out unrealized gains at time 2 |
| **Tax implication** | Incur capital gains tax | Capital gains tax depends on whether X makes a capital gain or capital loss | Defer capital gains tax to time 3 |

Under <u>strategy 1</u>, the taxable realization event of sale occurs at time 2. Under <u>strategy 2</u>, sale occurs at time 3. Both strategies call for an <u>ordinary sale</u>.

Under <u>strategy 3</u>, the taxpayer *simulates* the economic benefits of sale without selling the share at time 2, and *executes the actual sale* at time 3.[103] This calls for a <u>short sale</u>. The procedure for the tax deferral scheme in <u>strategy 3</u> is as follows:

---

[101] Lipton, Richard M, *New IRS Ruling Sanctions Some Variable Prepaid Forward Contracts*, 6 J. PASSTHROUGH ENTITIES 29 (2003)
[102] Halperin, Daniel, *Saving the Income Tax: An Agenda for Research*, 24 OHIO NUL REV. 493 (1998)
[103] Knoll, Michael S, *Regulatory Arbitrage Using Put-Call Parity*, 15 JOURNAL OF APPLIED FINANCE 1 (2005)

Time 1:

    Step 1.    X purchases 100,000 ABC shares at ₱50 per share.

Time 2:

    Step 2.    X borrows 100,000 ABC shares from a lender. He now has 200,000 ABC shares at hand: the <u>owned</u> and the <u>borrowed</u> ABC shares.

    Step 3.    X sells the 100,000 <u>borrowed</u> ABC shares to a third party buyer. X receives the sale proceeds at ₱100 per share.

Time 3:

    Step 4.    X replaces the 100,000 <u>borrowed</u> ABC shares by delivering 100,000 <u>owned</u> ABC shares to the lender.

The net effect of the transaction is that X cashes out the unrealized gain of ₱50 per share at time 2, and defers the accrual of capital gains tax to time 3.[104] This is illustrated as follows:

---

[104] *Supra* note 73.

## STRATEGY 3
## TAX DEFERRAL SCHEME

### CAPITAL LOSS[105] FROM SHARES OWNED

|  | Per Share (₱) | Total (₱) |
|---|---:|---:|
| Original Purchase Price (time 1) | 50 | 5,000,000.00 |
| Add: Unrealized Capital Gains (time 2) | 50 | 5,000,000.00 |
| Share Price (time 2) | 100 | 10,000,000.00 |
| Less: Unrealized Capital Loss (time 2) | 70 | 7,000,000.00 |
| Share Price (time 3) | 30 | 3,000,000.00 |
|  |  |  |
| Proceeds from Disposition of Shares (time 3)[106] | 30 | 3,000,000.00 |
| Less: Basis (time 1) | 50 | 5,000,000.00 |
| Capital Loss from Disposition of Owned Shares (time 3) | 20 | 2,000,000 |

### CAPITAL GAINS FROM SHORT SALE

|  | Per Share (₱) | Total (₱) |
|---|---:|---:|
| Proceeds from Sale of Borrowed Shares (time 2) | 100 | 10,000,000.00 |
| Less: Cost of Replacing Borrowed Shares (time 3) | 30 | 3,000,000.00 |
| Capital Gains from Short Sale (time 3) | 70 | 7,000,000.00 |

### NET CAPITAL GAINS COMPUTATION

|  | Per Share (₱) | Total (₱) |
|---|---:|---:|
| Capital Gains from Short Sale (time 3) | 70 | 7,000,000.00 |
| Less: Capital Loss from Disposition of Owned Shares (time 3) | 20 | 2,000,000.00 |
| Net Capital Gains (time 3) | 50 | 5,000,000.00 |
| Multiply by: Rate of Capital Gains Tax | 10% | 10% |
| Capital Gains Tax (time 3) | 5 | 500,000.00 |

COMPARISON OF STRATEGY 1 AND STRATEGY 3

Note that the amount of capital gains tax in Strategy 1 and Strategy 3 is the same, i.e. ₱5 per share. The only difference is the timing of accrual of tax: in Strategy 1, capital gains tax accrues at time 2, while in Strategy 3, capital gains tax accrues at time 3.

The second important difference between Strategy 1 and Strategy 3 is the timing of receipt of capital gains. In Strategy 1, X cashes out the capital gains of ₱50 per share *at the same time* that he becomes liable for capital gains tax of ₱5 per share. In Strategy 3, X cashes out the capital gains of ₱50 per share at time 2—i.e. ahead of the date he becomes liable for capital gains tax of ₱5 per share, which is at time 3.

The third difference is that Strategy 1 and Strategy 3 have different dates of realization. The date of realization dictates the date of accrual of capital gains tax. In Strategy 1, the date of realization is time 2. In Strategy 3, the date of realization is time 3. The reason Strategy 1's realization date is time 2 is that the sale was executed at time 2. And the reason X was able to defer the accrual of tax to time 3 is because the execution of sale was postponed to time 3.

---

[105] Notice that while the disposition of owned shares is at a loss, the short sale is at a gain.
[106] This is equivalent to the cost of replacing borrowed shares if X bought 100,000 ABC shares from the market instead of already owning them.

COMPARISON OF STRATEGY 2 AND STRATEGY 3

Under Strategy 2 and Strategy 3, shares were purchased at time 1 and held until time 3. Strategy 2 and Strategy 3 have the same original purchase price and the same unrealized capital loss as of time 3. However, in Strategy 2, X was exposed to the probability of capital loss. This is the reason why it incurred a capital loss of ₱20 per share at time 3. Had the share price at time 3 been higher than the original share price, X would have been exposed as well to the probability of improvements in capital gain. This did not transpire, but it could have.

In Strategy 3, it is true that X likewise held on to 100,000 ABC shares until time 3. However, X was already immune from the variability in the price movement of ABC shares between time 2 to time 3.

The drop in the share price from time 2 to time 3 did not eliminate the unrealized capital gains of X as of time 2. Had the share price increased from time 2 to time 3, X would not likewise experience an improvement in capital gains. This is the effect of the short sale transaction. The gain in the short sale neutralized the loss in the shares owned.

Accordingly, it was as if X "already sold" the appreciated shares at time 2, but became liable for capital gains tax only at time 3. Furthermore, the peso amount of his capital gains tax liability is based on the "realized" gains in time 2.

THE "ESTEE LAUDER" TRANSACTION

One of the most notorious cases on the tax deferral scheme we have thus far discussed is the Estee Lauder Transaction, involving the Estee Lauder Estate and the Estee Lauder Companies.[107] This led to the "constructive sale" rules in U.S. tax legislation, which sought to eliminate the tax avoidance opportunity created by short sale of identical securities.[108]

At time 1, the Estee Lauder Companies issued shares of stock to Estee Lauder. Estee Lauder, in turn, transferred her legal title over the shares to a trust agent called "EL 1994 Trust".[109]

At time 2, the Estee Lauder Co. issued its Initial Public Offering (IPO). Between time 1 and time 2, Estee Lauder's shares of stock have incurred significant unrealized gains. If the Trust sold the shares at time 2 (i.e. during the IPO), it would have been liable for millions in capital gains tax.[110]

*In order to "cash out" the unrealized gains at time 2 and defer the payment of capital gains tax to time 3*, the Trust borrowed shares of identical stock (i.e. Estee Lauder Co. shares) from Estee Lauder's own son, Leonard Lauder. The number of borrowed shares is the same as the number of shares held by the Trust. The Trust then

---

[107] *Supra* note 59.
[108] *Id.*
[109] *Id.*
[110] *Id.*

sold the borrowed shares in the market. The selling price is the prevailing market price during the IPO.[111]

With the short sale of Estee Lauder Co. shares, the Trust did not yet become liable for capital gains tax at time 2 because the realization event in a short sale only happens when the Trust finally replaces the borrowed shares. Secondly, the Trust continues to hold Estee Lauder Co. shares, which gains would be realized and taxed only when they are sold. Meanwhile, the Trust already obtained the proceeds of the short sale *in cash*.[112]

Moving forward from time 2, any subsequent gain in the owned Estee Lauder Co. shares would be offset by a loss in the short selling transaction, and *vice versa*. Therefore, the Trust is protected from an increase in capital gains tax as a result of increase in the fair market value of the owned shares.[113]

Only when the Trust decides to replace the borrowed Estee Lauder Co. shares — i.e. by delivering identical shares it already holds — will the Trust be liable for capital gains tax. And the tax would have the same amount had the Trust disposed the Estee Lauder Co. shares at time 2 without engaging in a short sale.[114]

This is a clear case of tax deferral. Incidentally, another son of Estee Lauder, Ronald Lauder, used the same technique. He borrowed 8.33 million shares from relatives and sold them short. One of the lenders was Ronald Lauder, his brother.[115]

EFFECT OF STEPPED-UP BASIS IN DEATH

The tax deferral scheme we have discussed thus far can evolve into a <u>complete tax avoidance scheme</u> if, between time 2 and time 3, death of the stockholder intervenes.[116] This is because, through death, the heir benefits from the <u>stepped-up basis</u> in properties acquired through inheritance.[117] Section 40(B)(2) of the NIRC states that the basis for determining gain or loss from sale or disposition of property shall be—

> The <u>fair market price or value</u> as of the <u>date of acquisition</u>, if the same was acquired by inheritance[.]

After time 2, but before time 3, the decedent stockholder passes title over ABC shares to the estate, and eventually to the heir.[118] The heir also inherits the contractual rights and obligations in the short sale, which remains open and unsettled after the decedent stockholder's death.[119] The heir then replaces the borrowed shares using the shares once owned by the decedent.[120] In disposing the owned shares, the <u>basis</u> of the disposition is no longer the <u>acquisition cost</u> of the shares, but the <u>fair market value</u> of the shares at the <u>time of death</u>.[121] The portion between the acquisition cost and fair market value becomes a realized gain, though <u>untaxed</u>.[122] The fair market value at the time of death is presumably close to the selling price of the shares, if the heir replaces the borrowed share by disposing the owned share within a short period after the decedent's death.[123]

---

[111] *Id.*
[112] *Id.*
[113] *Id.*
[114] *Id.*
[115] *Id.*
[116] Graetz, Michael J., *Taxation of Unrealized Gains at Death: An Evaluation of the Current Proposals*, VIRGINIA LAW REVIEW 830-859 (1973)
[117] Heckerling, Philip E, *The Death of the Stepped-Up Basis at Death*, 37 S. CAL. L. REV. 247 (1963)
[118] Article 774 of the New Civil Code states, "Succession is a mode of acquisition by virtue of which the property, rights and obligations to the extent of the value of the inheritance, of a person are transmitted through his death to another or others either by his will or by operation of law."
[119] Article 776 of the New Civil Code states, "The inheritance includes all the property, rights and obligations of a person which are not extinguished by his death."

Let us illustrate this tax avoidance scheme using the following prices of ABC shares:

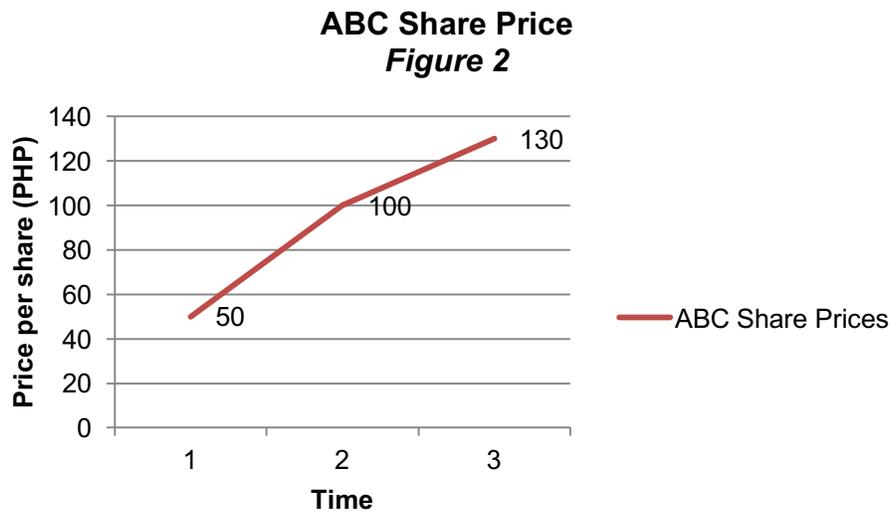

**ABC Share Price**
*Figure 2*

Assume that the stockholder dies between time 2 and time 3, at an intervening date where the price is already ₱130 per share.

The scheme proceeds as follows:

---

Time 1

    Step 1.    X purchases 100,000 ABC shares at ₱50 per share.

Time 2

    Step 2.    X borrows 100,000 ABC shares from a lender. He now has 200,000 ABC shares at hand: the <u>owned</u> and the <u>borrowed</u> ABC shares.

    Step 3.    X sells the 100,000 <u>borrowed</u> ABC shares to a third party buyer. X receives the sale proceeds at ₱100 per share.

Intervening date between Time 2 and Time 3

    Step 4.    X dies and transmits ownership of 100,000 ABC shares to Y, an heir, by operation of law. X also transmits contractual rights and obligations in the stock lending agreement to Y.

Time 3

    Step 5.    Y, the heir, replaces the 100,000 <u>borrowed</u> ABC shares by delivering 100,000 <u>owned</u> ABC shares to the lender.

Note that the intervention of death completely eliminates the capital gains tax on the realized gains,[124] illustrated as follows:

---

[124] Zelenak, Lawrence, *Taxing Gains at Death*, 46 VAND. L. REV. 361 (1993)

*TAX AVOIDANCE SCHEME*
*With Intervention of Death*

**CAPITAL GAINS FROM SHARES OWNED**

|  | Per Share (₱) | Total (₱) |
|---|---:|---:|
| Original Purchase Price (time 1) | 50 | 5,000,000.00 |
| Add: Unrealized Capital Gains (time 1-2) | 50 | 5,000,000.00 |
| Share Price (time 2) | 100 | 10,000,000.00 |
| Add: Unrealized Capital Gains (time 2-3) | 30 | 3,000,000.00 |
| Share Price (time 3) | 130 | 13,000,000.00 |
|  |  |  |
| Proceeds from Disposition of Shares (time 3) | 130 | 13,000,000.00 |
| *Less: Basis (intervening period between time 2 and time 3)* | ***130*** | ***13,000,000.00*** |
| Capital Gain from Disposition of Owned Shares (time 3) | 0 | 0 |

**CAPITAL LOSS FROM SHORT SALE**

|  | Per Share (₱) | Total (₱) |
|---|---:|---:|
| Proceeds from Sale of Borrowed Shares (time 2) | 100 | 10,000,000.00 |
| Less: Cost of Replacing Borrowed Shares (time 3) | 130 | 13,000,000.00 |
| Capital Loss from Short Sale (time 3) | 30 | 3,000,000.00 |

**NET CAPITAL LOSS COMPUTATION**

|  | Per Share (₱) | Total (₱) |
|---|---:|---:|
| Capital Gains from Disposition of Owned Shares (time 3) | 0 | 0.00 |
| Less: Capital Loss from Short Sale (time 3) | 30 | 3,000,000.00 |
| Net Capital Loss (time 3) | 30 | 3,000,000.00 |

A PROPOSAL FOR TAX REGULATION

Recall the steps of the tax deferral scheme discussed above:

(Step 1)   X <u>acquires</u> ABC share.
(Step 2)   X borrows an <u>additional</u> ABC share from a lender. He now has two ABC shares at hand: the <u>owned</u> ABC share and the <u>borrowed</u> ABC share.
(Step 3)   X sells the <u>borrowed</u> ABC share to a third party buyer.
(Step 4)   X replaces the <u>borrowed</u> ABC share *by delivering the <u>owned</u> ABC share* to the lender.

And recall how this is implemented using the price information and time periods in Figure 1, as follows:

Time 1:

Step 1.  X purchases 100,000 ABC shares at ₱50 per share.

Time 2:

Step 2.  X borrows 100,000 ABC shares from a lender. He now has 200,000 ABC shares at hand: the owned and the borrowed ABC shares.

Step 3.  X sells the 100,000 borrowed ABC shares to a third party buyer. X receives the sale proceeds at ₱100 per share.

Time 3:

Step 4.  X replaces the 100,000 borrowed ABC shares by delivering 100,000 owned ABC shares to the lender.

Under the present tax regime, there are two realization events, and both of these realization events happen only in step 4 at time 3. This is the reason why X can postpone the payment of capital gains tax to time 3, even though he had effectively "cashed out" the unrealized capital gains in step 3 at time 2. This is also the reason why X can benefit from the offsetting effect of the short sale.

The tax deferral scheme created by this situation can be addressed if we vary the timing of the two realization events. *It is submitted that one realization event should be recognized at time 2, and another realization event recognized at time 3*. We therefore propose the following language for tax regulation purposes:

> If a person owning property short sells an identical or substantially identical property, delivery by the short seller of the borrowed property to a third party buyer shall be deemed to be a disposition of the owned property, and the delivery of the owned property to replace the borrowed property shall be deemed to be a delivery of the borrowed property.

In the illustration, X owns ABC share and short sells an identical share. This fact should trigger the application of the proposed provision. Hence, where X short sells the borrowed ABC share to a third party buyer at time 2, such sale is construed for tax purposes to be an ordinary sale of the owned ABC share. Accordingly, X triggers the first realization event, and he becomes liable for capital gains tax at time 2 with respect to the capital gain in the *constructive* ordinary sale of the owned property. This is the meaning of the phrase, "delivery by the short seller of the borrowed property to a third party buyer shall be deemed to be a disposition of the owned property".

Considering that the short sale of the borrowed share has been constructively treated as an ordinary sale of the owned share, X can no longer replace the borrowed share at time 3 using the owned share. Hence, delivery of the owned share to replace the borrowed share should be constructively treated as delivery of the borrowed share to the lender. This is the second realization event. This is the meaning of the phrase, "delivery of the owned property to replace the borrowed property shall be deemed to be a delivery of the borrowed property."

In *Ocier vs. Commissioner of Internal Revenue*[125], Jerry Ocier transferred 4.9 million shares of Best World Resources Corporation (hereinafter referred to as BW shares) to Dante Tan. The transfer was allegedly made pursuant to a stock lending agreement, denominated as a trust declaration, with Ocier as lender and Tan as borrower. The BIR construed the transfer as a sale and assessed a deficiency capital gains tax of ₱17.86 million to be paid by Ocier. Disregarding the claim of Ocier that the transfer was made pursuant to a stock lending agreement, the Court of Tax Appeals (CTA) states that a securities borrowing and lending agreement is a non-taxable transaction, but only if it complies with the formalities required by regulation. In this case, the trust declaration between Ocier and Tan was not prepared in accordance with the BIR guidelines on securities borrowing and lending agreements. Accordingly, Ocier was liable for deficiency capital gains tax.

The case is illustrative of the existing regulatory regime on the borrowing of shares (which necessarily applies to short sales of stock). However, it only addresses one aspect of a short selling transaction: the transfer of shares between lender and borrower. Recall that there are two agreements involved in a short sale: the borrowing of securities and the short sale proper, where the short seller disposes the borrowed stock to a third party, who is not privy to the stock lending transaction.

If we assume that the trust declaration transferring the stock to Tan is for the purpose of selling the borrowed stock, Tan has an obligation to deliver 4.9 BW shares back to Ocier in the future, in order to replace the borrowed BW shares. In *People vs. Tan*[126], Dante Tan was prosecuted for concealing beneficial ownership of 84 million BW shares, which are identical to the allegedly borrowed shares of stock in *Ocier*. Under the present tax treatment of short sales, Tan would not be liable for capital gains tax if he short sells the 4.9 borrowed BW shares. He would only be liable when he replaces 4.9 BW shares to Ocier. Under the proposed tax treatment of short sales, however, any disposition to be made in short selling BW shares, when Tan currently owns 84 million BW shares, should be construed as a sale of a portion of the 84 million BW shares, such that any profit would be deemed as a taxable realized gain.

Under the proposed tax realization rule, we modify the treatment of the tax deferral scheme through short sale of stock, as follows:

(Step 1)  X acquires ABC share.
(Step 2)  X borrows acquires an additional ABC share from a lender. He now has two ABC shares at hand: the owned ABC share and the borrowed ABC share.
(Step 3)  X sells the ~~borrowed~~ owned ABC share to a third party buyer.
(Step 4)  X replaces the borrowed ABC share by delivering the ~~owned~~ **borrowed** ABC share to the lender.

Note the stricken off terms, which pertain to the original tax treatment, and their replacements, representing the new tax treatment under our proposal. Step 3 triggers the first realization event, and step 4 triggers the second realization event. This is implemented using Figure 1 prices, as follows (again, note the stricken off terms and their replacements):

---

[125] CTA Case No. 6831, 02 February 2009
[126] G.R. No. 167526, July 26, 2010

Time 1:

    Step 1.    X purchases 100,000 ABC shares at ₱50 per share.

Time 2:

    Step 2.    X borrows 100,000 ABC shares from a lender. He now has 200,000 ABC shares at hand: the <u>owned</u> and the <u>borrowed</u> ABC shares.

    Step 3.    X sells the 100,000 ~~borrowed~~ **owned** ABC shares to a third party buyer. X receives the sale proceeds at ₱100 per share.

Time 3:

    Step 4.    X replaces the 100,000 <u>borrowed</u> ABC shares by delivering 100,000 ~~owned~~ **borrowed** ABC shares to the lender.

The first realization event happens at time 2 and the second happens at time 3. With this modified timing of realization events, there is no more opportunity for tax deferral using short sale of stock. At time 2, X incurs a capital gain and pays a capital gains tax, as follows:

|  | Per Share (₱) | Total (₱) |
|---|---|---|
| Selling Price (time 2) | 100 | 10000000 |
| Less: Acquisition Cost (time 1) | 50 | 5000000 |
| Capital Gain (time 2) | 50 | 5000000 |
| Multiply by: CGT rate | 10% | 10% |
| Capital Gains Tax (time 2) | 5 | 500000 |

At time 3, X incurs another taxable capital gain and pays capital gains tax, as follows:

|  | Per Share (₱) | Total (₱) |
|---|---|---|
| Proceeds from Short Sale (time 2) | 100 | 10,000,000 |
| Less: Cost of Replacement of Borrowed Share (time 3) | 30 | 3,000,000 |
| Capital Gain (time 3) | 70 | 7,000,000 |
| Multiply by: CGT rate | 10% | 10% |
| Capital Gains Tax (time 3) | 7 | 700,000 |

If we compare the existing tax rule with the proposed tax rule, the existing rule yields a *deferred* capital gains tax of ₱5 per share, which is only collectible at time 3. Under the proposed tax rule, a total capital gains tax of ₱12 per share is due to the National Government: ₱5 per share is collectible at time 2 and ₱7 per share is collectible at time 3. On top of that, we have eliminated the tax deferral opportunity in the income tax system.

Short sales are not illicit *per se*. These activities are expressly allowed in the Securities Regulation Code (R.A. No. 8799) and the National Internal Revenue Code

(R.A. No. 8424). Short sales also serve the crucial function of providing liquidity in the market for securities.[127] For instance, it helps brokers address failed trades and intraday defaults.[128]

Moreover, short selling of securities identical to those already owned by the short seller is not an automatic indicator of a dubious motivation to defer or avoid capital gains tax liability.[129] Investors holding highly appreciated financial positions in shares of stock may wish to retain stock ownership for the purpose of perpetuating corporate control and other incidents of corporate ownership, while insulating themselves from financial market risks.[130] Second, traders who are expecting an abnormal but temporary downturn in stock prices can enter short sales as a hedging strategy to prevent unexpected capital losses.[131]

The prevalence of short selling activities, however, can have a material adverse impact on the BIR collection performance if left unregulated. The BIR collection performance for years 2010 to 2014 shows a 68.7% increase in capital gains tax collection.[132] The actual and target collection figures[133] are summarized as follows:

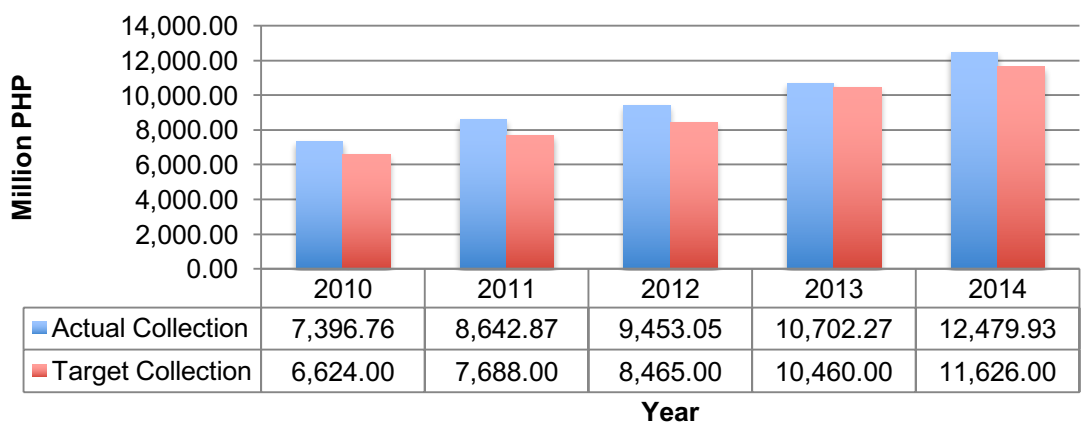

**Capital Gains Tax Collection**

| Year | 2010 | 2011 | 2012 | 2013 | 2014 |
|---|---|---|---|---|---|
| Actual Collection | 7,396.76 | 8,642.87 | 9,453.05 | 10,702.27 | 12,479.93 |
| Target Collection | 6,624.00 | 7,688.00 | 8,465.00 | 10,460.00 | 11,626.00 |

(Million PHP)

The increasing trend in actual and target collection for capital gains tax reflects, in part[134], the fact that the PSE is one of the "fastest growing bourses in southeast Asia."[135] There are at present 264 listed companies and 132 active trading participants.[136] Between late 1990s and early 2000s, there were less than 5,000 trades per day.[137] By 2013, the average daily trades reached 30,000.[138] As of March 2016, total market capitalization reached ₱13.95 trillion, up from ₱8.87 trillion in 2010.[139] Presently, the average daily value turnover is ₱6.78 billion, which was only at ₱4.95 billion in 2010.[140]

---

[127] *Supra* note 21.
[128] *Id.*
[129] Ulcickas, Simon D., *Internal Revenue Code Section 1259: A Legitimate Foundation for Taxing Short Sales Against the Box or a Mere Makeover?* 39 WILLIAM & MARRY LAW REVIEW 4, 1355 (1998)
[130] Hu, Henry TC, and Bernard Black, *Empty voting and hidden (morphable) ownership: Taxonomy, implications, and reforms*, THE BUSINESS LAWYER 1011-1070 (2006)
[131] *Supra* note 129.
[132] BIR Annual Tax Statistics for CY 2010-2014, *available at*: http://www.bir.gov.ph/index.php/transparency/bir-annual-report.html
[133] *Id.*
[134] Capital gains tax collection, as reported in the BIR Annual Tax Statistics, includes the disposition of all capital assets, including real estate and securities.
[135] *Supra* note 21.

      Under our proposal, the taxpayer owning a share of stock must recognize a gain upon entering into a short sale of an identical stock. The proposal intends to curb the use of short selling transactions to implement tax deferral and avoidance strategies. It would be an irony to see a more advanced capital market coupled with underperforming capital gains tax collection efforts by the BIR. Surely, the regulatory institutions did not intend to erode National Government revenues by institutionalizing short sale of securities in the Philippines.

---

[136] *See* The Philippine Stock Market 1st Quarter 2016 Briefer, *available at:* http://pse.com.ph/resource/1Q16%20v4.pdf

[137] Crisostomo, Regina Georgia R., Sarah L. Padilla, and Mark Frederick V. Visda, *Philippine Stock Market in Perspective*, 12th National Convention on Statistics, *available at*: http://www.nscb.gov.ph/ncs/12thncs/papers/INVITED/IPS-21%20Finance%20Statistics/IPS-21_3_Philippine%20Stock%20Market%20in%20Perspective.pdf

[138] *Id.*

[139] *Supra* note 136.

[140] *Id.*